\documentclass[german,english]{bvm2020}

\def\articlenumber{2828}

\date{}
\title{Deep OCT Angiography Image Generation for Motion Artifact Suppression}

\titlerunning{Deep OCTA Motion Artifact Suppression}
\author{Julian~Hossbach$^{1,2}$, Lennart~Husvogt$^{1,3}$, Martin~F.~Kraus$^{1,4}$, James~G.~Fujimoto$^3$, Andreas~K.~Maier$^1$}

\authorrunning{Hossbach et al.}

\institute{%
$^1$ Pattern Recognition Lab, Friedrich-Alexander-Universit\"at Erlangen-N\"urnberg\\
$^2$ SAOT, Friedrich-Alexander-Universit\"at Erlangen, Germany\\
$^3$ Biomedical Optical Imaging and Biophotonics Group, MIT, Cambridge, USA\\
$^4$ Siemens Healthineers AG, Erlangen}

\email{julian.hossbach@fau.de}

\begin{document}

%
\selectlanguage{english}

\maketitle

\begin{abstract}
Eye movements, blinking and other motion during the acquisition of optical coherence tomography (OCT) can lead to artifacts, when processed to OCT angiography (OCTA) images. Affected scans emerge as high intensity (white) or missing (black) regions, resulting in lost information. The aim of this research is to fill these gaps using a deep generative model for OCT to OCTA image translation relying on a single intact OCT scan. Therefore, a U-Net is trained to extract the angiographic information from OCT patches. At inference, a detection algorithm finds outlier OCTA scans based on their surroundings, which are then replaced by the trained network. We show that generative models can augment the missing scans. The augmented volumes could then be used for 3-D segmentation or increase the diagnostic value.
\end{abstract}

\section{Introduction}
Optical coherence tomography (OCT) enables a cross-sectional and non-invasive imaging of the eye's fundus in axial direction and thus, by depicting the retina's anatomical layers, results in a high diagnostic value in ophthalmology. A further significant increase for the examination is offered by the OCT angiography (OCTA), which exhibits the microvasculature of the retina and choroid. For example, microaneurysms, caused by diabetic retinopathy, can be identified in OCTA images \cite{2828-01}.

One popular, solely image based, OCTA imaging technique relies on the varying reflectance of the flowing red blood cells over time, while the signal of surrounding tissue is almost stationary. By decorrelating 2 or more OCT scans, acquired over a short period of time at the very same location, vessels are enhanced and form an OCTA scan \cite{2828-02}. While the acquisition of a single OCT image, also called B-scan, is rapidly done by combining 1-D axial scans in lateral direction, a stack of B-scans is considerably slower.

As OCTA is derived from image changes over successively acquired B-scans, other origins of image differences than blood flow can result in artifacts. One cause of such artifacts is motion, which leads to a misalignment of the B-scans of consecutive volumes. The origin of the motion can be rooted in pulsation, tremor and microsaccades, involuntary executed by the subject. Misalignment can lead to high decorrelation values and manifest as white stripes in en-face projections (along axial direction) of the OCTA volume \cite{2828-03}. Another source of artifacts, showing as black lines in the projection, is blinking, which results into lost OCT scans and therefore no OCTA image at that location.

More elaborated OCTA algorithms, like the Split-spectrum Amplitude Decorrelation Angiography, can compensate for motion in the axial direction but still relies on several aligned structural B-scans and quarters the axial resolution \cite{2828-04}. A recent algorithm consists of a Deep Learning (DL) pipeline for the extraction of angiographic information from a stack of OCT B-scans \cite{2828-05}. As both methods rely on matching B-scans, they are prone to the aforementioned artifacts. We propose a new generative DL based strategy to replace detected erroneous OCTA B-scans using an image-to-image translation architecture from a single OCT scan. Exhibiting these blind spots of conventional algorithms can increase the usability for physicians and improve further processing like vessel segmentation.

\section{Materials and methods}

\subsection{Defect scan detection}
\label{2828-detection}
The first step to replace defect OCTA scans is the detection algorithm, which triggers the OCT-OCTA translation network for identified angiographic scans.

Threshold based detection is commonly used, thus we adapt this approach to include the aforementioned artifacts. The summed flow signal of OCTA scans is calculated for each B-scan and compared to an upper and lower value. Instead of global thresholds, local limits $\theta_{\rm l/u}$ are determined with $\theta_{\rm l} = \mu_{\rm 16}-\tau_{\rm l}\sigma_{\rm 16}$ and $\theta_{\rm u} = \mu_{\rm 5}+\tau_{\rm u}\sigma_{\rm 5}$ for the lower and upper bound, respectively. The mean $\mu_x$ and the variance $\sigma_x$ is calculated over $x-1$ nearest B-scans and the scan itself. For the lower bound, a broader neighborhood is considered, as well as the parameter $\tau_{\rm l}=0.029$ was selected less strict as this threshold only filters blank scans. With $\tau_{\rm u}=0.0255$ and a higher sensitivity to local changes, $\theta_{\rm u}$ detects motion corrupted scans and outliers.

\subsection{Data set and processing}
The next step we process our data such that it can be used for the OCT-OCTA translation network.
The data for this project consists of 106 volumetric OCTA scans, which were each generated from 2 structural volumes. The volume size is 500 $\times$ 500 $\times$ 465/433 (number of B-scans, lateral and axial) pixels over an area of 3 $\times$ 3\ts mm or 6 $\times$ 6\ts mm for 54 and 52 volumes, respectively. The data is split into a training, validation and test data sets of 91, 15 and 10 volumes and zero padded to a mutual axial size of 480 pixels to compensate for the varying dimensionality. Each B-scan is normalized to the range of 0 to 1 using the global maximum value.

The information on vessels in OCT scans is likely not depended on the global B-scan, but rather locally encoded. Therefore, we can crop the B-scans along the lateral axis into patches of size 480 $\times$ 128\ts pixels to reduce the input size for the neural network (NN). During training we randomly select 100 patches per OCT-OCTA B-scan pair, which passed the detection algorithm, to augment the data. At inference, the input B-scan is cropped into 5 overlapping patches, such that the first and last 8 columns can be removed at the composition of the outputs.

Patches where the depicted structures were cropped by the top or bottom image margins due to the curvature of the retina were removed.

\subsection{OCT-OCTA translation}
After the detection and outlier rejection, the remaining data is used to train a supervised NN for image-to-image translation.

\begin{figure}[t]
    \centering
    \caption{Layout of the utilized U-Net for image translation with Dense Blocks and average pooling of 2 $\times$ 2 in the encoding part. The decoder combines nearest neighbor interpolation and residual blocks for upsampling.}
    \includegraphics[width=0.98\textwidth]{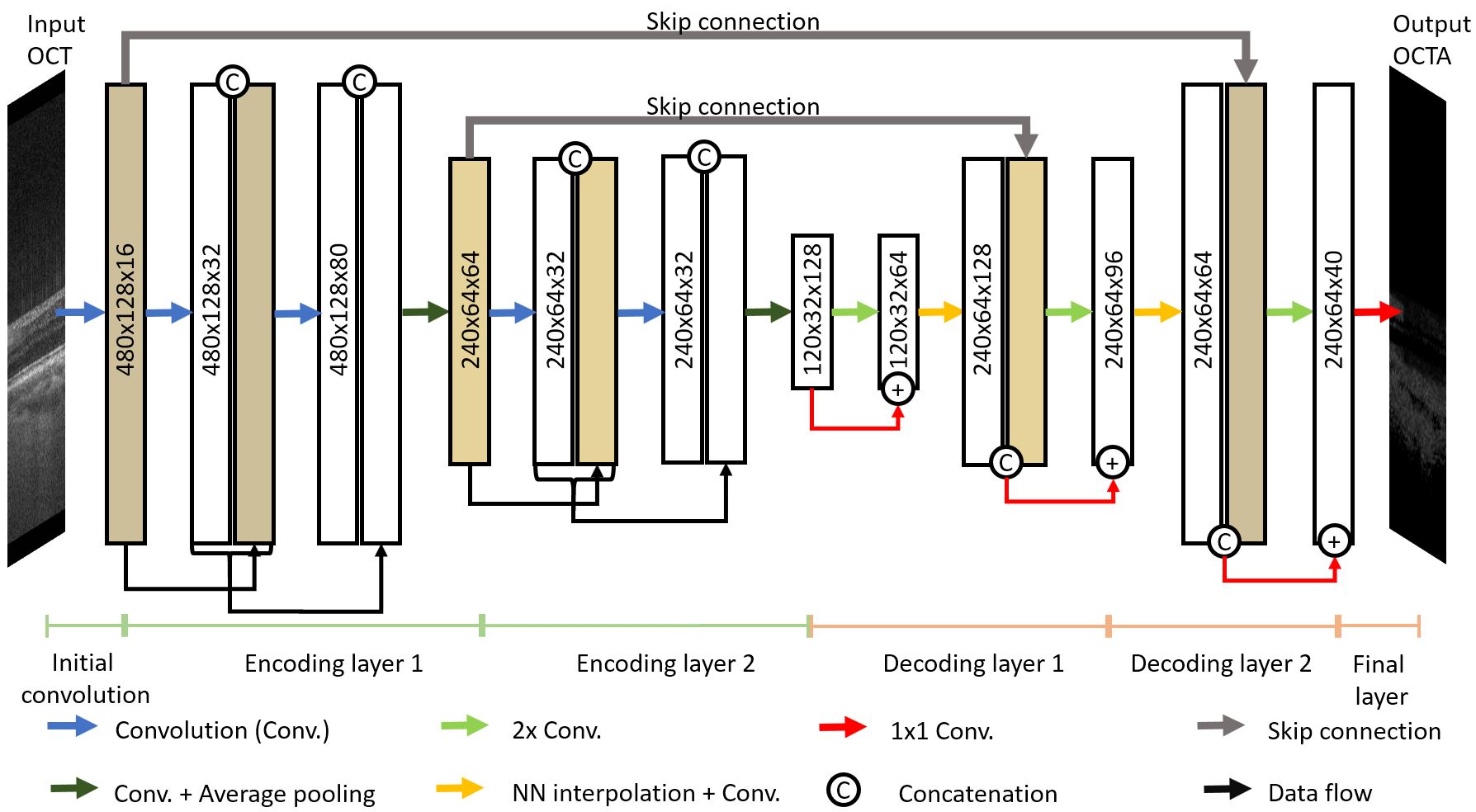}
    \label{2828-fig:unet}
\end{figure}

To achieve such domain transform, U-Nets \cite{2828-06} has been shown to be suitable. The layout of the developed U-Net is illustrated in Fig. \ref{2828-fig:unet}. In the encoding section, after an initial convolution, two blocks are used to each halve the spatial input resolution. Each block consists of 2 dense blocks, which concatenates the block's input with its output along the channel direction after a convolution as input for the next layer. The transition block learns a 1 $\times$ 1 convolution and reduces the width and height by average pooling (2 $\times$ 2). The latent space is decoded by 2 decoding layers with an upsampling operation and final 1 $\times$ 1 convolution with a sigmoid activation. Each decoding layer is built from a residual layer followed by a nearest neighbor interpolation (not in the last layer) and a convolution to avoid upsampling artifacts. If not stated otherwise, the kernel size is 3 $\times$ 3 pixels, and every convolution is followed by leaky ReLU (slope 0.1) as non-linearity and batch normalization.

The training process minimized the L2 norm between the generated patch and the ground truth. For a few training epochs, a further preprocessing step smooths the target OCTA images with a 3 $\times$ 3 $\times$ 3 median filter to reduce the speckle noise in the target data and to enforce a higher focus on vessel information. In addition, from related work, it can be expected that speckle noise is also reduced in DL generated OCTA scans \cite{2828-05}.

\section{Results}
The evaluation is 3 folded. First, we evaluate the generated OCTA B-scans from the test data by comparing them with the ground truth at intact locations. Secondly, we look at the en-face projection of the fake angiographic volume and its reference. Due to the anatomical structure of the choroid, the amount of blood vessels leads to noisy OCTA signals without hardly any recognizable individual vessels in the ground truth. In projections of our OCTA data, this leads to a reduced contrast of the vessels in upper layers. Therefore, we segment the vitreous as upper bound and the inner plexiform layer as lower bound using the OCTSEG-tool \cite{2828-07} of two volumes and only use the area in between for the evaluation of the projection. Finally, the detected defect scans are replaced and evaluated.

\begin{figure}[H]
    \centering
    \caption{The first four images show the input, target, output and absolute error for one example B-scan. The projections of the segmented volumes are shown below in order input, target and completely generated output. Note the artifacts are removed in the generated projection.}
    \includegraphics[width=0.98\textwidth]{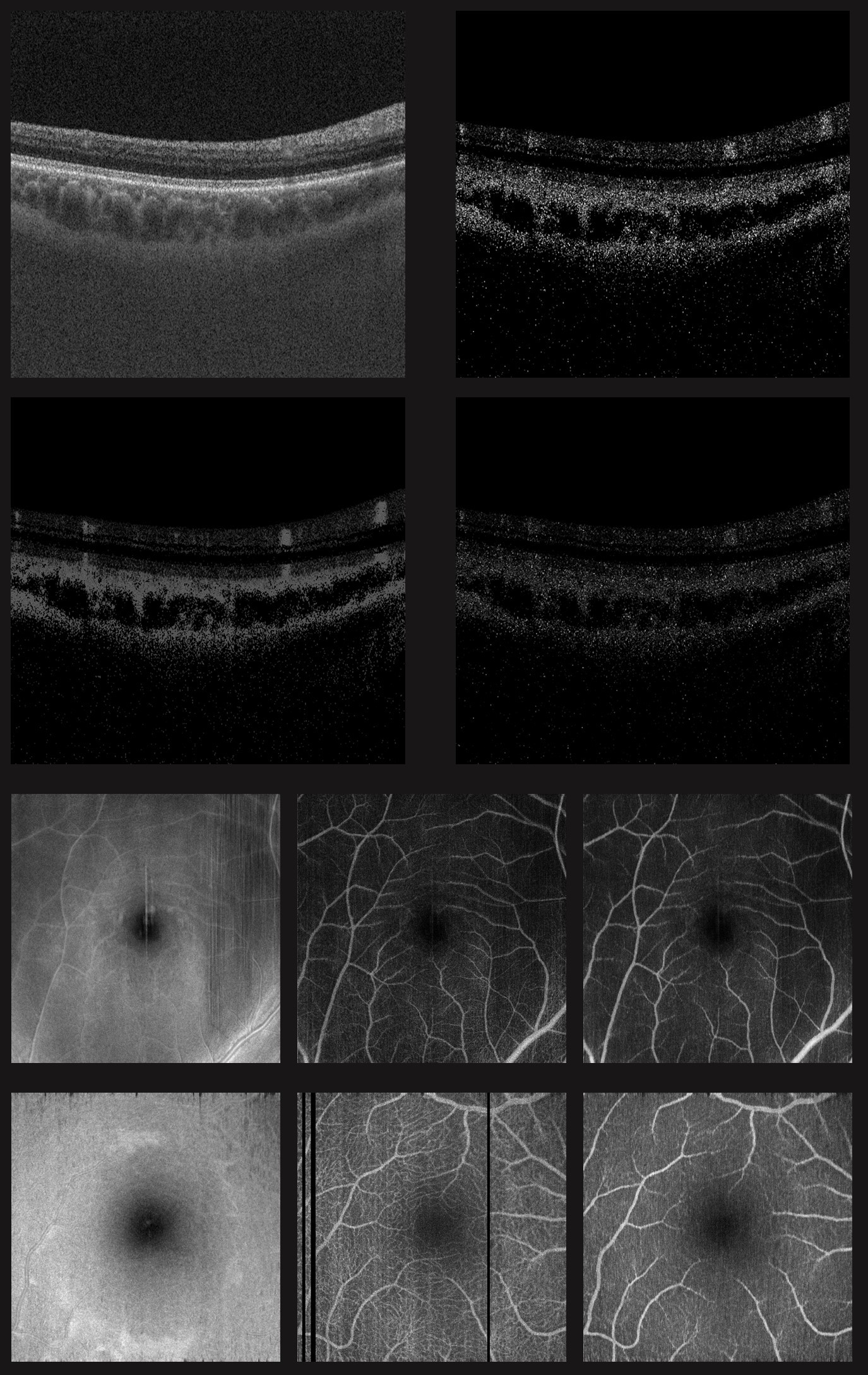}
    \label{2828-fig:images}
\end{figure}

The first row of Tab. \ref{2828-tab:bscan} shows the mean absolute error (MAE), mean square error (MSE) and the structural similarity (SSIM) between the generated scan and the ground truth, averaged over all intact test B-scans. Example B-scans are depicted in Fig. \ref{2828-fig:images}. The top left image shows the input OCT scan, to its right, the target angiographic image is shown. In the row below, the output image and the absolute difference to the ground truth are displayed. As expected from Liu et al. \cite{2828-05}, the result has reduced noise structure.

\begin{SCtable}[5][htb]
    \begin{tabular}{cccc}
        \hline
          & MAE & MSE & SSIM \\ \hline
          B-scans: & 0.019 & 0.002 & 0.761 \\ \hline
          Seg. proj. 1: & 0.0096 & 0.0001 & 0.9978 \\ \hline
          Seg. proj. 2: & 0.0138 & 0.0002 & 0.9896 \\ \hline 
          Unseg. proj.: & 0.0096 & 0.0001 & 0.93 \\ \hline
    \end{tabular}
    \caption{The table shows the mean absolute error (MAE), mean squared error (MSE) and the structural similarity (SSIM) for all test B-scans, the 2 segmented projections and the remaining unsegmented projections.}
    \label{2828-tab:bscan}
\end{SCtable}

The image metrics of the segmented projections are written in Tab. \ref{2828-tab:bscan} in the second and third row, the metric of the unsegmented volumes in the last row. In the 3rd and 4th row of Fig. \ref{2828-fig:images}, the segmented projection of the OCT, target and output volume are exhibited. Regarding also the less noisy structure, it can be concluded that not only speckle noise is diminished, but also smaller vessels. Nonetheless, all major vessels are extracted accurately. 

\begin{SCfigure}[5][tb]
	\includegraphics[width=0.25\textwidth]{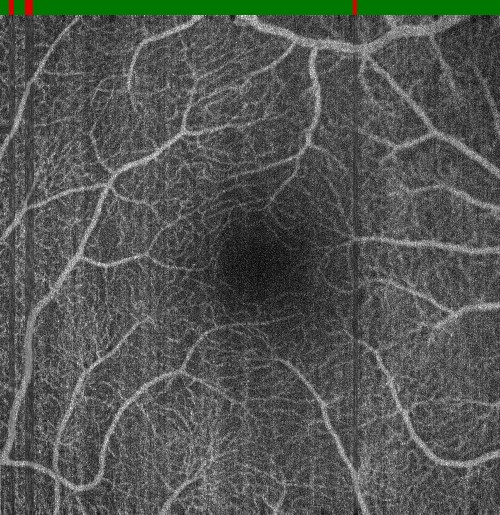}
    \caption{The defect scans of the projection in Fig. \ref{2828-fig:images} were replace with the generated ones. On top, the line marks the detected defect (red) and intact (green) scans.}
    \label{2828-fig:replace}
\end{SCfigure}

The target projection of the second example in Fig. \ref{2828-fig:images} clearly shows several artifacts as a result of blinking. In Fig. \ref{2828-fig:replace}, we used the lower threshold to detect the defect scans and merge the network's result into the detected gaps, which are marked with the red stripe on top.

\section{Discussion and conclusion}
Based on a single intact OCT scan, the NN can approximate useful OCTA images. Major vessels have a sharper edge and seemingly a better contrast. Smaller vessels are not extracted accurately, but this might be rooted in the fact, that a single structural scan is insufficient for such details. As a result, inserted scans can be identified, when compared to the neighborhood. Furthermore, our approach still needs a correct input B-scan, otherwise the artifacts will remain.

As the requirements for our approach only comprise a single scan at locations with artifacts and the network with its preprocessing, an integration into the OCTA imaging workflow is easily possible and can improve the image quality easily.


\ack
We gratefully acknowledge the support of NVIDIA Corporation with the donation of the Titan Xp GPU used for this research.

\bibliographystyle{bvm2020}

\bibliography{2828}
\marginpar{\color{white}E\articlenumber} 
\end{document}